# Providing educational accountability for Local Authorities based upon sampling pupils within schools: moving away from simplistic school league tables.

by

Harvey Goldstein, University of Bristol

George Leckie, University of Bristol

Lucy Prior, University of Bristol

Corresponding author:

Harvey Goldstein

School of Education

35 Berkeley Square

Bristol BS8 1JA

h.goldstein@bristol.ac.uk

## Abstract
This paper proposes an alternative educational accountability system for England that moves away from simplistic comparisons, or league tables, among schools towards a more nuanced reporting at a level of education authorities or other school groupings. Based upon the sampling of pupils within schools, it proposes the use of quantitative pupil assessment data as screening devices that provide evidence within an integrated accountability system that includes inspection. At the same time, it aims to provide a richer set of data than currently available for research as well as accountability purposes. We argue that if carefully implemented within a context of school improvement, such a system has the potential to largely eliminate the deleterious side effects and curriculum distortions of the present system. While being proposed within the context of the current English secondary school system, the proposals will have relevance for other phases of schooling and similar systems in other countries.

# Introduction

In a number of educational systems, and since the late 1980s in England, schools have been ranked and publicly compared on the basis of pupil attainment measures. With the advent of the UK Department for Education (DfE) National Pupil Database in England in 2002 (DfE 2019) this has become a routine exercise where Key Stage 2 (KS2) and Key Stage 4 (KS4) scores and exam results have been used. In the latter case KS2 attainment scores have also been used to 'adjust' KS4 exam results in an attempt to provide a 'fairer' comparison. A detailed review of past and present school performance measures in England including a discussion of alternative procedures can be found in Leckie and Goldstein (2017).

During the period when such 'league tables' have become common, there has also been a realisation that there are considerable deleterious side effects associated with such accountability systems (NAHT, 2018) and there have been suggestions about how they could be modified or abandoned (Leckie and Goldstein, 2019). Thus, for example, Yang et al (1999) proposed a system where school comparisons were not published but discussed with schools in a context of 'improvement' rather than 'punishment' or 'shaming'. This concept was further elaborated in Foley and Goldstein (2012) who also considered health, criminal justice, and other public sector areas where institutions are routinely monitored and held to account using data.

An important feature of debates around this issue, is that policymakers of most persuasions have appeared unwilling to remove the comparative performance aspect of league tables, so that caveats and evidence of their unfitness for purpose have been largely ignored. More recently, Ofsted and others have started to downplay the role of school comparisons based on DfE numerical summaries and are now instead concentrating on independently using both internal and external reviews to assist management and help to adjust curriculum and other procedures (Guardian, 2018; OFSTED, 2018).

In this paper we propose an approach that has the promise of removing the most deleterious effects of comparative league tables, while providing a numerically based tool that can be used as a sensitive screening instrument alongside more detailed inspection systems. In the next section we outline the structure of such a system, and this is followed by a series of analyses using existing data to explore its feasibility.

# Sampling based accountability and research

In the spirit of Foley and Goldstein (2012) we define the principal role of numerical estimates of institutional performance based upon pupil attainments as providing summary statistics that can be used as screening instruments to highlight apparent exceptionally poor or exceptionally good institutional or group achievements. The idea is that these are not to be used to make definitive judgements, but rather to identify where further examination is likely to be profitable.  For such a purpose it is neither necessary nor desirable to produce ranked comparisons among all such groups or institutions, and our proposed procedures are designed in such a way that they are in fact not suitable for constructing public league tables.

The basic idea is that pupil attainments are only collected from a subsample of children. We wish to collect enough data for each Local Authority (LA) so that a chosen percentage of LAs with the most extreme scores will have 'value added' estimates that are significantly different from average and can be investigated further. Likewise, a percentage of the most outlying schools may also have significant value-added estimates thus allowing those schools to be studied if that is required. For the majority of schools, however, since only a proportion of pupils will be selected, league tables become infeasible. Furthermore, if accountability is directed solely at the larger units of LAs, we can also encrypt (pseudonymise) the school identifiers in order to make their true identities more difficult to determine.

We need to determine the amount of data from each LA and the example given below demonstrates results from selecting 250 pupils at random with equal probability from each of the 151 LAs with mainstream schools in England. We also explore the consequences of a second possible sampling strategy whereby for each LA we randomly select an equal number of pupils from each school to achieve a sample of approximately 250 from each LA. It remains a further research issue as to how many pupils should be selected and in what fashion, and the present analyses can be viewed as a pilot exercise for this. We should point out that, although we have chosen to discuss LAs as our principal unit of interest, our arguments can be applied to other groupings, such as academy chains, regional groups, etc.

We note that at the same time as using the data for accountability purposes, the overall sample of more than 35,000 pupils will be large enough to conduct sensitive research that explores longitudinal relationships. We will illustrate some of these possibilities.

## Data

We use the DfE NPD to recreate the underlying pupil-level Attainment 8 and KS2 score dataset for the cohort of pupils who took their GCSE examinations in 2016. Attainment 8 is essentially a total score across eight subjects (DfE, 2018) that was introduced in 2016 and serves as the KS4 attainment outcome. These data are fully described in Leckie and Goldstein (2019) and so we only summarise their main features here. The dataset contains a range of standard pupil background and school characteristics. Pupil characteristics include: age, gender, ethnicity, language (whether they speak English as an additional language), SEN (special educational needs status), FSM (eligible for free school meals at some time in the preceding 6 years: an indicator of poverty) and a deprivation measure of the pupil's residential neighbourhood as proxied by the Income Deprivation Affecting Children Index (IDACI) for their own residential postcode (MHCLG, 2015). School characteristics include: region, type, admissions policy (selective or not), age range, gender mix, religious denomination and IDACI score of the school neighbourhood. The final analysis sample consists of 502,851 pupils in 3,098 schools located in 151 local authorities across the nine regions of England.

In the first set of analyses we use a simple random sample of 250 pupils from each LA. This would be justified on the basis that it avoids estimates for large LAs being more accurate than for smaller LAs. This yields 37,454 pupils with 1% (332) having missing information on ethnic status which will be excluded for modelling. To aid interpretation, we normalise pupils' Attainment 8 scores so that it is measured in standard deviation units and we centre the KS2 score on its approximate mean value of 4. The DFE categorizes the total KS2 score into 34 groups. However, a plot of the mean Attainment 8 score by group suggests a quadratic relationship (see Leckie and Goldstein, 2019, Figure S2), and this relationship is used in all analyses to fit the mean trend. Table 1 shows the results of a 3-level model of normalised Attainment 8 score, fitting main effects for the above pupil characteristics. In addition, we have fitted school level effects, namely the denominational status of the school, the gender composition, school IDACI and type of admissions policy. The categories used for all covariates are given in Table 1. All multilevel models are fitted using MLwiN version 3.04 (Charlton et al, 2019).

| Table 1. Attainment 8 as outcome by selected covariates for the simple random sample within LAs of 250 pupils. Pupils are nested within schools within Local Authorities. Attainment 8 scores have been normalised. KS2 score has been centred around a value of 4. Pupils=37,122, Schools=2,277, LAs=151. | | |
|---|---|---|
| **Covariate** | **Estimate** | **Standard error** |
| Intercept | -0.560 | 0.014 |
| KS2 score - 4 | 0.786 | 0.006 |
| (KS2 score – 4) squared | 0.258 | 0.004 |
| Girl pupil | 0.159 | 0.007 |
| Ethnic: Black | 0.129 | 0.014 |
| Ethnic: Asian | 0.114 | 0.013 |
| Ethnic: Other | 0.011 | 0.013 |
| Language of home not English | 0.257 | 0.011 |
| IDACI for pupil's residence | -0.527 | 0.026 |
| Free school meals eligible | -0.231 | 0.007 |
| SEN support | -0.248 | 0.010 |
| SEN statement | -0.376 | 0.022 |
| School admission: Grammar | 0.302 | 0.026 |
| School admission: Secondary Modern | -0.023 | 0.028 |
| IDACI for pupil's school | -0.171 | 0.039 |
| School gender: boys | 0.044 | 0.021 |
| School gender: girls | 0.020 | 0.018 |
| School denomination: Church of England | 0.024 | 0.019 |
| School denomination: Roman Catholic | 0.069 | 0.015 |
| School denomination: Other Christian | 0.004 | 0.031 |
| School denomination: Jewish | 0.401 | 0.084 |
| School denomination: Muslim | 0.270 | 0.114 |
| School denomination: Sikh | -0.094 | 0.241 |
| ***Random effects*** | | |
| Intercept: LA | 0.003 | 0.001 |
| Intercept: school | 0.026 | 0.002 |
| Intercept: pupil | 0.330 | 0.002 |
| -2*loglikelihood | 65936.3 | |
| | | |
| Reference categories for discrete covariates are as follows: Gender – boy; Ethnic – White; language of home – English; Free school meals – not eligible; SEN – no SEN status, School admissions type – comprehensive; school gender – mixed; school denomination - none | | |

For comparison we fit the same model to the full dataset in Table 2. As we would expect the results are essentially the same, but with reduced standard errors. Since in both models we have fitted both school and LA random effects, our estimates are essentially within-school-within LA ones so the disproportionate sampling of LAs as in Table 1 will not bias the estimates relative to these full cohort ones which similarly are within-school-within-LA estimates.

| Covariate | Estimate | Standard error |
|---|---|---|
| **Table 2. Attainment 8 as outcome by selected covariates for the full dataset. Pupils within schools within Local Authorities. Attainment 8 scores have been normalised. KS2 score has been centred around a value of 4. Pupils = 498,608, Schools = 3,097, LAs=151*.** | | |
| Intercept | -0.617 | 0.009 |
| KS2 score - 4 | 0.796 | 0.002 |
| (KS2 score – 4) squared | 0.247 | 0.001 |
| Girl pupil | 0.160 | 0.002 |
| Ethnic: Black | 0.119 | 0.004 |
| Ethnic: Asian | 0.111 | 0.004 |
| Ethnic: Other | 0.086 | 0.004 |
| Language of home not English | 0.245 | 0.003 |
| IDACI for pupil's residence | -0.543 | 0.008 |
| Free school meals eligible | -0.219 | 0.002 |
| SEN support | -0.239 | 0.003 |
| SEN statement | -0.346 | 0.006 |
| Grammar | 0.313 | 0.019 |
| Secondary Modern | -0.030 | 0.021 |
| IDACI for pupil's school | -0.193 | 0.029 |
| School gender: boys | 0.080 | 0.018 |
| School gender: girls | 0.044 | 0.015 |
| School denomination: Church of England | 0.045 | 0.015 |
| School denomination: Roman Catholic | 0.069 | 0.012 |
| School denomination: Other Christian | 0.061 | 0.023 |
| School denomination: Jewish | 0.248 | 0.063 |
| School denomination: Muslim | 0.242 | 0.070 |
| School denomination: Sikh | -0.031 | 0.188 |
| *Random effects* | | |
| Intercept: LA | 0.004 | 0.001 |
| Intercept: school | 0.032 | 0.001 |
| Intercept: pupil | 0.327 | 0.001 |
| -2*loglikelihood | 866677.2 | |
| Reference categories for discrete covariates are as follows: Gender – boy; Ethnic – White; language of home – English; Free school meals – not eligible; SEN – no SEN status, School admissions type – comprehensive; school gender – mixed; school denomination – none | | |

Most of the coefficients for pupil and school characteristics in Table 1 are in line with previous research. We note the very small amount of variance accounted for by the LA (0.2%) compared to 1.3% for a null model with no covariates (not shown). Of some interest is the significantly better performance of boys in boy' schools compared to mixed schools (0.44 SD units). Roman Catholic, Jewish and Muslim schools do better than non-denomination schools. The IDACI score for where the pupil lives is also significant with a higher score (and therefore a relatively more deprived area) associated with lower Attainment 8 and there is a smaller but also negative association for the IDACI score associated with the location of the school.

We also note that some of the coefficients do appear to differ somewhat between Tables 1 and 2. We have not studied this in detail, but it appears that this is largely due to associations between LA and these covariates and between school and these covariates. If we wish to make marginal inferences about the whole population a weighted analysis would be appropriate (Rabe-Hesketh and Skrondal, 2006).

In recognition of the potential importance of differential effectiveness for educational accountability, the notion that the influence of schools and LAs on student learning may vary across different pupil groups (see for example, Strand, 2016), we have also studied some random coefficients and the results are displayed in Table 3.

Table 3. Attainment 8 as outcome by selected covariates for the simple random sample. Pupils within schools within Local Authorities. Attainment scores have been normalised. KS2 score has been centred around a value of 4. Pupils=37,122, Schools=2,277, LAs=151.

| Covariate | Estimate | Standard error |
| --- | --- | --- |
| Intercept | -0.558 | 0.014 |
| KS2 score - 4 | 0.787 | 0.007 |
| (KS2 score – 4) squared | 0.257 | 0.005 |
| Girl pupil | 0.160 | 0.006 |
| Ethnic: Black | 0.127 | 0.014 |
| Ethnic: Asian | 0.112 | 0.013 |
| Ethnic: Other | 0.109 | 0.013 |
| Language of home not English | 0.255 | 0.011 |
| IDACI for pupil's residence | -0.543 | 0.008 |
| Free school meals eligible | -0.233 | 0.008 |
| SEN support | -0.247 | 0.010 |
| SEN statement | -0.376 | 0.029 |
| Grammar | 0.301 | 0.027 |
| Secondary Modern | -0.028 | 0.028 |
| IDACI for pupil's school | -0.129 | 0.029 |
| School gender: boys | 0.042 | 0.021 |
| School gender: girls | 0.020 | 0.018 |
| School denomination: Church of England | 0.016 | 0.019 |
| School denomination: Roman Catholic | 0.071 | 0.014 |

| School denomination: Other Christian | 0.010 | 0.031 |
|---|---|---|
| School denomination: Jewish | 0.392 | 0.083 |
| School denomination: Muslim | 0.296 | 0.111 |
| School denomination: Sikh | -0.096 | 0.228 |
| *Random effects* | | |
| **LA**: Variance Intercept | 0.004 | 0.001 |
| Covariance Intercept: KS2 score | -0.001 | 0.001 |
| Variance KS2 score | 0.002 | 0.001 |
| **School:** Variance Intercept | 0.018 | 0.002 |
| Covariance Intercept: KS2 score | 0.000 | 0.001 |
| Variance KS2 score | 0.005 | 0.001 |
| Covariance Intercept: FSM | 0.004 | 0.002 |
| Covariance Intercept: KS2 score | 0.001 | 0.002 |
| Variance FSM | 0.014 | 0.003 |
| Covariance Intercept: SEN statement | 0.001 | 0.007 |
| Covariance KS2 score: SEN statement | 0.003 | 0.005 |
| Covariance FSM: SEN statement | -0.002 | 0.008 |
| Variance SEN statement | 0.171 | 0.027 |
| **Pupil; intercept** | 0.322 | 0.003 |
| -2*loglikelihood | 65716.9 | |
| | | |
| Reference categories for discrete covariates are as follows: Gender – boy; Ethnic – White; language of home – English; Free school meals – not eligible; SEN – no SEN status, School admissions type – comprehensive; school gender – mixed; school denomination - none | | |

As shown in Table 3, there are significant random coefficients across schools for the KS2 score, pupils with a SEN statement, and those eligible for free school meals. Thus, the relationships between Attainment 8 and these covariates varies across schools. This in turn implies that schools do not have constant positive or negative effects on their students, rather the magnitude of the effects they have varies across these pupil groups. At the LA level it is just the KS2 score that has a random coefficient for these covariates, as shown.

We now illustrate, for LAs, how many have estimated residuals (value added estimates) that are significantly different from average based on both 95% and 90% confidence intervals. Figure 1a shows this for the model in Table 3, with 95% intervals for the LA random intercept (i.e. at a KS2 value of 4.0). Figure 1b shows the corresponding plot for 90% intervals. Likewise Figures 2a and 2b show the corresponding intervals for the estimated LA random coefficient of KS2 score.

**Figure 1a. Estimated LA 95% confidence intervals for intercept.**

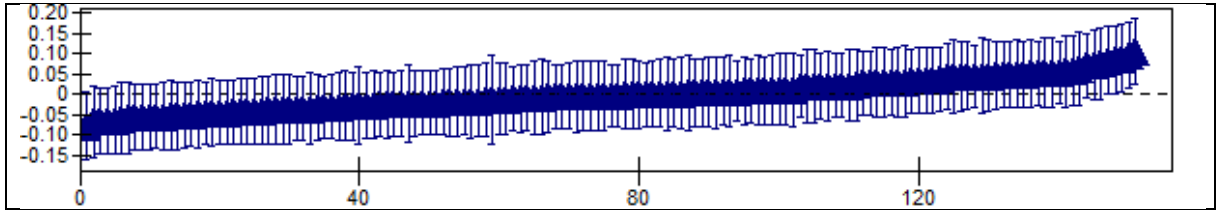

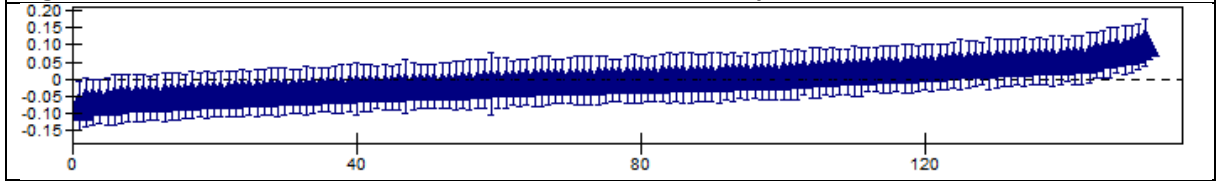

**Figure 1b. Estimated LA 90% confidence intervals for intercept.**

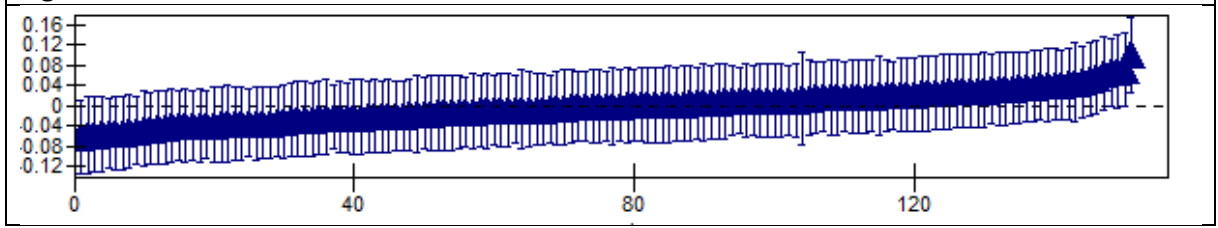

**Figure 2a. Estimated LA 95% confidence intervals for the random coefficient of KS2 score.**

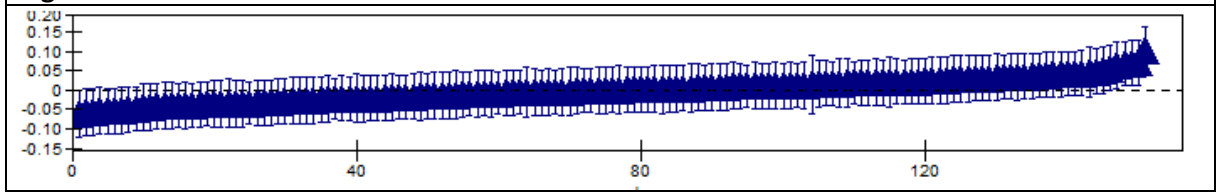

**Figure 2b. Estimated LA 90% confidence intervals for the random coefficient of KS2 score**

We see that for the 95% LA intervals there are only 1 or 2 intervals significantly higher than zero and none significantly lower than zero. In the spirit of using the data as a screening device, in order to obtain more LAs for further exploration we could simply choose a shorter confidence interval. Thus for example, for 90% intervals there are 6 significantly above the mean of the residuals (zero) for the pupils having the mean KS2 score with none significantly below. For the coefficient of KS2 there are 5 significantly higher than zero and none significantly less. The choice of one or more thresholds for screening at different levels of intensity, will depend partly on resources available, and an assessment of both false positive and false negative rates. We shall return to this issue in the discussion. In comparison with the model that utilises the full data set, these intervals are on average three times the length, so that quite a lot of precision is lost through a sampling approach, as can be seen in the estimated standard errors in Tables 1 and 2. This will generally be of more concern for research data analysis than for accountability purposes.

We can also study other random coefficients. For example, looking at schools, for those pupils with a SEN statement less than 1% (20 schools) have 95% intervals) significantly higher than zero with just 1 school significantly less.

## Sampling equal numbers of pupils per school

It may be felt desirable that each school within an LA should supply equal numbers of pupils to avoid the situation where a simple random sample is selected within a LA and thus where large schools are overrepresented in the sample and thus will provide more accurate value added estimates for small schools. Indeed, under the previous sampling criterion 26% primarily small schools (820) were not sampled at all.

If, instead, we sample an equal number of pupils from every school in the country, one problem is that when making an inference which is representative for any given LA, which is the primary target of inference in this paper, we would need to weight by the numbers in each school. For modelling purposes, unless there are important relationships between school size and the response, conditional on covariates, we will obtain consistent estimates for parameters. Nevertheless, it does in fact appear to be the case, as noted below, that such relationships do exist. This would therefore seem to make this particular sampling scheme somewhat problematic.

We would not, therefore, wish to pursue this particular sampling procedure in general but for completeness we present just the model results, as given in Table 4.

| Table 4. Attainment 8 as outcome by selected covariates for equal numbers of pupils within schools within each LA. Pupils within schools within Local Authorities. Attainment 8 scores have been normalised. KS2 score has been centred around a value of 4. Pupils=37,357, schools=3,095, LAs=150. | | |
|---|---|---|
| Covariate | Estimate | Standard error |
| Intercept | -0.614 | 0.016 |
| KS2 score - 4 | 0.767 | 0.007 |
| (KS2 score – 4) squared | 0.250 | 0.004 |
| Girl pupil | 0.161 | 0.006 |
| Ethnic: Black | 0.112 | 0.014 |
| Ethnic: Asian | 0.115 | 0.013 |
| Ethnic: Other | 0.105 | 0.013 |
| Language of home not English | 0.265 | 0.011 |
| IDACI for pupil's residence | -0.524 | 0.026 |
| Free school meals eligible | -0.211 | 0.008 |
| SEN support | -0.244 | 0.010 |
| SEN statement | -0.344 | 0.028 |
| Grammar | 0.306 | 0.027 |
| Secondary Modern | -0.046 | 0.028 |
| IDACI for pupil's school | -0.170 | 0.039 |
| School gender: boys | 0.097 | 0.023 |
| School gender: girls | 0.044 | 0.020 |
| School denomination: Church of England | 0.010 | 0.019 |
| School denomination: Roman Catholic | 0.065 | 0.015 |
| School denomination: Other Christian | 0.037 | 0.031 |
| School denomination: Jewish | 0.162 | 0.077 |
| School denomination: Muslim | 0.260 | 0.090 |
| School denomination: Sikh | -0.148 | 0.235 |
| ***Random effects*** | | |
| **LA**: | | |
| Variance Intercept | 0.004 | 0.001 |
| Covariance Intercept: KS2 Score | -0.001 | 0.001 |
| Variance KS2 score | 0.002 | 0.001 |
| **School:** | | |
| Variance Intercept | 0.022 | 0.002 |
| Covariance Intercept: KS2 score | 0.001 | 0.001 |
| Variance KS2 score | 0.006 | 0.002 |
| Covariance Intercept: FSM | 0.004 | 0.002 |
| Covariance Intercept: KS2 score | 0.001 | 0.001 |
| Variance FSM | 0.014 | 0.003 |
| Covariance Intercept: SEN statement | 0.000 | 0.008 |
| Covariance KS2 score: SEN statement | 0.024 | 0.006 |
| Covariance FSM: SEN statement | -0.003 | 0.002 |
| Variance SEN statement | 0.208 | 0.029 |
| **Pupil; intercept** | 0.314 | 0.003 |
| -2*loglikelihood | 65866.3 | |
| | | |
| Reference categories for discrete covariates are as follows: Gender – boy; Ethnic – White; language of home – English; Free school meals – not eligible; SEN – no SEN status, School admissions type – comprehensive; school gender – mixed; school denomination - none | | |

The results set out in Table 4 are similar to those in Table 3. We note, for example, that the variation between schools in the mean difference for those with a SEN statement and those without SEN, is greater where we have sampled equal numbers in each school as compared to a simple random sample of pupils within LA, equivalent to schools sampled proportional to size. This suggests that the larger the school the greater the difference, and this would be an area for further research.

## Discussion and recommendations.

We have illustrated how a random sample of equal numbers of pupils from each LA allows inferences about pupil attainment relationships to be explored as well as providing information about accountability. Selecting equal numbers from each LA, in the context of accountability, is consistent with judging each LA on the same basis, or as having equal weights within an accountability system. In terms of our model inferences, these estimate 'within LA' as well as 'within school' relationships, since school and LA random effects are included in the model. We note, however, that LAs are different sizes, so that, if our model is correct, our inferences are consistent, but if there are unmodelled factors associated with the response (for example the size of the LA itself) and the LA size then to obtain population unbiased estimates we would need to carry out a weighted analysis, weighting by LA size. We do not explore this here, but this could be a feature of further research. We have chosen a sample size of approximately 250 from each LA. As we have seen this does allow us to carry out detailed analyses, including random coefficients in order to explore whether LAs have differential effects on different pupil groups. Nevertheless, a topic for further research and piloting prior to the implementation of any system is exploring appropriate sample size, consistent with not placing an undue burden on schools and LAs while providing adequate numbers for nuanced data analyses.

In the case where we sampled equal numbers of pupils within each school, this might be an appropriate strategy if we also wished to provide an element of accountability for schools where equal weight was attached to each school. From a research perspective this would imply additional weighting to obtain population consistent inferences. Since, however, school level accountability is not the main purpose of our analysis we shall not pursue this. We note, however, that Table 4 which is the same model that is given in Table 3 but based on the sample arising from this second sampling scheme, provides somewhat different estimates. This is essentially because Table 3 is based upon a self-weighting sample within each LA and so provides LA consistent estimates, whereas the sample modelled in for Table 4 does not. This implies that there are likely to be other factors that are associated with the response and school characteristics that have not been included.

The implementation of a system such as we have described would need careful piloting and we now discuss what we believe are essential components of such a system.

First of all, as we have assumed, only a sample of pupils would be chosen each year for KS2 testing and subsequent follow up. Given movement between schools and losses to the population, we would need to sample rather more than our analyses above imply, and this would need to be ascertained through piloting. We do not propose that the NPD should be discontinued. On the contrary, it would continue to provide a framework for monitoring the educational system, with linkages to other datasets, and in particular would act as the sampling frame as described below. We appreciate that reformulating the structure of the NPD would require careful thought, in particular to enable continuity and a smooth transition.

The number of pupils per LA will also determine the interval estimate size for each LA. This in turn will determine how many conventional 95% or 90% intervals do not overlap the mean, and these are typically chosen as those that should be followed up using, for example, inspectors, either to understand if there are particular problems or policies responsible for a better or worse than average performance. For comparison some inspections could also take place within schools not exhibiting any extreme behaviours. Nevertheless, basing a follow up choice on a criterion directly related to the size of the LA sample, seems generally undesirable. A more rational approach would involve a function of available resources and a knowledge of the false positive and false negative rates given particular sample sizes and different follow up threshold values. Experiments based upon different sample sizes and threshold choices would be an area for further research. It would also seem desirable to allocate follow up resources in relation to the probability, say, of detecting a problem where one actually exists (See, for example, Goldstein, 1972).

One issue that requires explicit attention is that of privacy and data confidentiality. The transparency of the system needs to be strengthened and in particular, parents and guardians of individual pupils need to be supplied with information about the uses of the data and also with options to opt out of supplying certain kinds of information, including the freedom to remove test scores, or to request that they not be used for particular purposes (Defend Digital Me, 2019). In this respect, it will be important to make a distinction between use for 'research' purposes and for other purposes, such as commercial ones. Most importantly, the option to opt out of participation should be available for any use of data that leads to the result being publicly available; which is the essence of any research application.

We will refer to the modified NPD system as NPD2. To obtain further assessment data, for example relating to the KS1 or KS2 or KS3 curriculum stages then NPD2 would provide the 'sampling frame' from which a suitable sample of pupils would be assessed and it would be those assessments that would be used to provide value added (or other research relevant data). Thus these assessments plus other data from NPD2 (such as demographic information) would be available for analysis as we have demonstrated above. The main advantage of such a system is that the assessments (testing) would be carried out for specific (but possibly regularly recurring) purposes and that only a sample of pupils from any one school would be chosen – consistent with being small enough to make most school-level comparisons infeasible. Currently, for example, DfE chooses a threshold of 10 such that school cohorts below this size do not get reported and this might serve as an initial rule of thumb, although larger samples would also be acceptable and possibly necessary for LAs with small numbers of schools. At the same time, the sample selected should be adequate to provide useful LA data for screening purposes, as discussed above. As mentioned earlier, we can provide further protection against the construction of school league tables that might be derived from released datasets by encrypting or pseudonymising the school identifiers.

Suppose that we wanted to track a set (possibly all) LAs in terms of value added over the secondary period through to 18 years of age. As opposed to the way in which current test score data are collected, we could also use a matrix sampling design (Goldstein and James, 1983) to avoid too heavy a burden on each pupil, with the advantage that this would allow us to collect a rich set of data. This is a common design and, for example, is used in the Programme for International Student Assessment, PISA, (OECD, 2001) where each pupil takes only a randomly chosen subset of items from a large number designed to capture many aspects of curricula. The group could be followed up at various times, but in any case their public exam records would be available from NPD2 so that value added estimates could be obtained. If, for illustration, we aimed to obtain 250 pupils in each LA then this would result in a sample of average size 25 if there were 10 schools in the LA, with different ways of sampling within schools as we have illustrated. Such a sample, since it would be national would also provide useful information about the system as a whole, and have the ability to use the NPD2 as a frame so that, importantly, longitudinal relationships can be studied, for example, with respect to pupil characteristics such as eligibility for FSM, that change over time. One of the advantages of such a system is that different samples of pupils can be selected for different purposes, including one-off studies to gather information.

NPD2 would satisfy a number of demands currently being made to deal with issues of teacher burden and anxiety. Any additional assessments would be done using external teams (awarded contracts to do so) and without involving undue teacher time. Consultation and piloting in different contexts would of course be essential as would careful consideration of how a transition from NPD to NPD2 is handled. Participation in collecting such additional information would of course require (opt out) consent and non-response would be present, but since the target population and their basic characteristics are known, there is a whole variety of methods now available that can help substantially with adjusting for any biases that arise from lack of consent or other reasons for selected pupils being missing from the data.

Any proposal to implement a scheme of this kind should pay attention to any initiatives currently happening, and the possibility of linking data with other datasets, especially in health, and we note that as of 2019 there are several initiatives looking at the construction of total population 'spines' that can act as devices for linking datasets (see for example Statslife, 2018).

By removing pressure from schools to compete within the current accountability system there will be a need to consider how a different, and superior, 'accountability for improvement' system at the school level might be encouraged. There is a strong argument for considering this as separate from any assessment but utilising LA data and between-school data (see below) as part of an inspection framework.

There would be many advantages of acquiring information about interactions between schools with LAs. These might be use of shared facilities, movements of pupils for specialist purposes, or staff movements. Such data could be a component of the models fitted to the data and also useful to a revised inspection system that was not totally centred on individual schools but concerned with the local environment within which a school operated, especially how it interacted with other schools and the LA. It is envisaged that such an emphasis on group-level activities, both in terms of the group level nature of assessments and within-group interactions, would encourage cooperation among schools rather than competition.

A publicly accountable national organisation, perhaps somewhat analogous to the Office of Qualifications and Examinations Regulation (OFQUAL), and independent of government ministers, could assume responsibility for all of the design and analysis tasks and be staffed by suitably qualified experts and other representatives of interest groups, including some of those concerned with protecting privacy, as well as learned societies.

It would be important to pilot any new system carefully, in perhaps a small number of LAs, especially with regard to the features of any transition period that would phase out league tables as currently provided and move towards the use of value added data as screening instruments allied to a sensitive inspection mechanism developed within an overall accountability system motivated by an overall concern for school improvement.

## Acknowledgements

This research was funded by UK Economic and Social Research Council grant ES/R010285/1. We are grateful to Professor Gemma Moss for helpful comments.

# References

Charlton, C., Rasbash, J., Browne, W.J., Healy, M. and Cameron, B. (2019) MlwiN Version 3.04 (University of Bristol, Centre for Multilevel Modelling).

Defend Digital Me (2019). National Pupil Database. https://defenddigitalme.com/national-pupil-database/ (accessed 5.7.2019).

DfE (2018) Secondary accountability measures: Guide for maintained secondary schools, academies and free schools (London, Department for Education).

DfE (2019) Collection: National pupil database (London, Department for Education). https://www.gov.uk/government/collections/national-pupil-database (accessed 5.7.2019).

Foley, B. & Goldstein, H. (2012) Measuring success: League tables in the public sector (London, British Academy).

Goldstein H & James AN. (1983). Efficient Estimation for a Multiple Matrix Sample Design.. British J. of Math. & Stat. Psych. **36** 167-174.

Goldstein H. (1972). The Allocation of Resources in Population Screening. A decision theory model. . Biometrics **28** 499-518

Guardian (2018) Ofsted inspectors to stop using exam results as key mark of success. (11/10/2018): https://www.theguardian.com/education/2018/oct/11/ofsted-to-ditch-using-exam-results-as-mark-of-success-amanda-spielman. (accessed 5.7.2019)

Labour (2019) Jeremy Corbyn's speech at NEU conference: https://labour.org.uk/press/jeremy-corbyns-speech-neu-conference/ (accessed 5.7.2019).

Leckie, G. and Goldstein, H. (2019). The importance of adjusting for pupil background in school value-added models: A study of Progress 8 and school accountability in England. British Educational Research Journal. DOI: Vol. 45, No. 3, June 2019, pp. 518–537. 10.1002/berj.3511

Leckie, G., & Goldstein, H. (2017). The evolution of school league tables in England 1992-2016: 'contextual value-added', 'expected progress' and 'progress 8'. British Educational Research Journal, 43, 193-212. DOI: 10.1002/berj.3264.

MHCLG (2015) National Statistics: English indices of deprivation 2015 (London, Ministry of Housing, Communities & Local Government): https://www.gov.uk/government/statistics/english-indices-of-deprivation-2015 (accessed 5.7.2019)

NAHT (2018) Improving school accountability (London, National Association of Head Teachers).

OECD (2001) Knowledge and skills for life: first results from Programme for International Student Assessment (Paris, OECD).

Ofsted (2018). Press release: Chief Inspector sets out vision for new Education Inspection Framework (11/10/2018): https://www.gov.uk/government/news/chief-inspector-sets-out-vision-for-new-education-inspection-framework (accessed, 5.7.2019).

Rabe‐Hesketh, S. & Skrondal, A. (2006) Multilevel modelling of complex survey data. *Journal of the Royal Statistical Society: Series A (Statistics in Society)*, 169, 805-827.

Statslife (2018). https://www.statslife.org.uk/news/3805-longitudinal-studies-need-strong-use-and-impact-evidence-base (accessed 5.7.201).

Strand, S. (2016) Do some schools narrow the gap? Differential school effectiveness revisited, *Review of Education*, 4, 107–144